# Frustrated magnetism of the $S$ = 1 Trillium-Lattice Oxide $Li_2NiGe_3O_8$

Yuya Haraguchi*
*Department of Applied Physics and Chemical Engineering, Tokyo University of Agriculture and Technology, Koganei, Tokyo 184-8588, Japan*
*Corresponding author: chiyuya3@go.tuat.ac.jp

We report magnetization and heat-capacity measurements on the ordered-spinel oxide $Li_2NiGe_3O_8$, where $Ni^{2+}$ ions with $S$ = 1 form a single three-dimensional trillium lattice. Powder x-ray diffraction confirms a cubic ordered-spinel structure with space group $P4_132$ or $P4_332$. The inverse susceptibility $H/M$ follows Curie-Weiss behavior above 50 K with an effective magnetic moment $\mu_{eff}$ = 3.124(4) $\mu_B$ per Ni and a Weiss temperature $\theta_W$ = –0.21(1) K, but deviates smoothly below about 10 K. The magnetic heat capacity $C_{mag}/T$ shows a broad maximum near 3 K with a wide tail to about 10 K, and the entropy recovered between 2 and 40 K is about 88% of $R$ ln 3. The broad heat-capacity maximum is compared with Monte Carlo results for the local ferromagnetic Ising model on the trillium lattice using a characteristic scale $J$ of order 7 K, while the inverse susceptibility shows only qualitative similarity to the theoretical curve. These results establish $Li_2NiGe_3O_8$ as a rare $S$ = 1 single-trillium oxide with frustrated magnetic correlations. The present data provide an experimental platform for discussing the relation between Heisenberg-like and spin-ice-like regimes on the trillium lattice.

## I. Introduction

Geometrical frustration in magnets often produces states in which strong interactions coexist with the absence of simple long-range order [1-5]. Corner-sharing triangle and tetrahedron networks are representative examples, leading to kagome antiferromagnets, pyrochlore spin ice, and related cooperative paramagnets [1-5]. In pyrochlore spin ice, local moments constrained along local 111 directions obey a two-in two-out rule on each tetrahedron, generating a macroscopically degenerate manifold with finite residual entropy [5-8]. The same class of materials also established the broader ideas of magnetic monopoles, Coulomb phases, and quantum descendants of classical spin ice [9-13]. These developments motivate the search for related constrained states on other three-dimensional frustrated lattices.

The trillium lattice is a three-dimensional chiral network of corner-sharing equilateral triangles that was originally identified as a high-symmetry crystal net in chiral structures [14]. Theoretically, it supports several distinct frustrated limits. For nearest-neighbour Heisenberg interactions, classical calculations revealed an extended cooperative paramagnetic regime, partial-order physics, and eventual low-temperature ordering [15-17]. For the local-axis ferromagnetic Ising model studied by Redpath and Hopkinson, each triangle satisfies a two-out one-in or one-out two-in rule, producing a broad soft peak in the heat capacity, a finite residual entropy, and a shoulder-like inverse susceptibility without conventional long-range order [18]. These features make the trillium lattice a natural three-dimensional analogue in which Heisenberg-type and spin-ice-type frustration can be compared within the same geometry.

Historically, most experimentally known trillium-lattice magnets were metallic. MnSi and related B20 compounds provided the first prominent platform, revealing helimagnetism, fluctuating chiral order, and skyrmion-lattice physics on the trillium Mn sublattice [19,20]. Other metallic or intermetallic trillium systems, including EuPtSi, EuPtGe, and CeIrSi, showed that the same geometry can host broad fluctuation regimes, first-order transitions, and unusual chiral magnetic phases [21-27]. These materials established the importance of the trillium network, but the coexistence of itinerancy, conduction electrons, or complex rare-earth physics complicates direct comparison with simple local-moment spin models.

A major recent development is the emergence of genuinely insulating trillium magnets with localized moments. $K_2Ni_2(SO_4)_3$ was identified as a system of two coupled $S$ = 1 trillium lattices and was reported to host a highly correlated dynamic state at zero field and a field-induced quantum spin-liquid regime [28,29]. $Na[Mn(HCOO)_3]$ provided the first experimental realization of the nearest-neighbour Heisenberg antiferromagnet on a single trillium lattice, with strong frustration and classical spin-liquid behavior above its ordering temperature [30]. This family has rapidly expanded to include additional phosphate-based trillium systems, such as $KSrFe_2(PO_4)_3$, $K_2CrTi(PO_4)_3$, and $KBaCr_2(PO_4)_3$, which show strong frustration, broad correlated regimes, or multiple field-dependent anomalies on single or double trillium networks [31-33]. Taken together, these studies have established the Heisenberg-dominated side of insulating trillium-lattice magnetism.

What remains much less explored experimentally is not simply a new trillium magnet, but a localized-moment single-trillium material whose bulk thermodynamics can be meaningfully compared with the Ising-like limiting model on the same lattice. In this context, the presence of $Ni^{2+}$ alone does not guarantee spin-ice physics. For $S$ = 1 $Ni^{2+}$, ideal octahedral coordination is often close to isotropic, whereas substantial easy-axis or easy-plane anisotropy can emerge only through details of the ligand field and the associated zero-field splitting [34-38]. The key question is therefore not whether $Li_2NiGe_3O_8$ contains $Ni^{2+}$, but how its observed thermodynamics should be positioned relative to the Heisenberg-like and spin-ice-like limiting models on the trillium lattice.

$Li_2NiGe_3O_8$ is an attractive candidate in this regard. Previous structural and transport studies showed that $Li_2NiGe_3O_8$ crystallizes in an ordered-spinel framework in which the next-nearest-neighbour $Ni^{2+}$ sublattice forms the trillium network [39-41]. Because $Li_2NiGe_3O_8$ is an insulator and $Ni^{2+}$ carries localized $S$ = 1 moments in octahedral coordination, it offers an opportunity to examine whether spin-ice-like correlations can develop in a single oxide trillium lattice.

The central point of the present work is that $Li_2NiGe_3O_8$ exhibits broad magnetic correlations on the $S$ = 1 trillium lattice. The inverse susceptibility starts to deviate from Curie-Weiss behavior below about 10 K even though $\theta_W$ is nearly zero, and the magnetic heat capacity shows a broad maximum around 3 K rather than a sharp lambda-type anomaly. We compare the broad heat-capacity anomaly with Monte Carlo results for the local ferromagnetic Ising model on the trillium

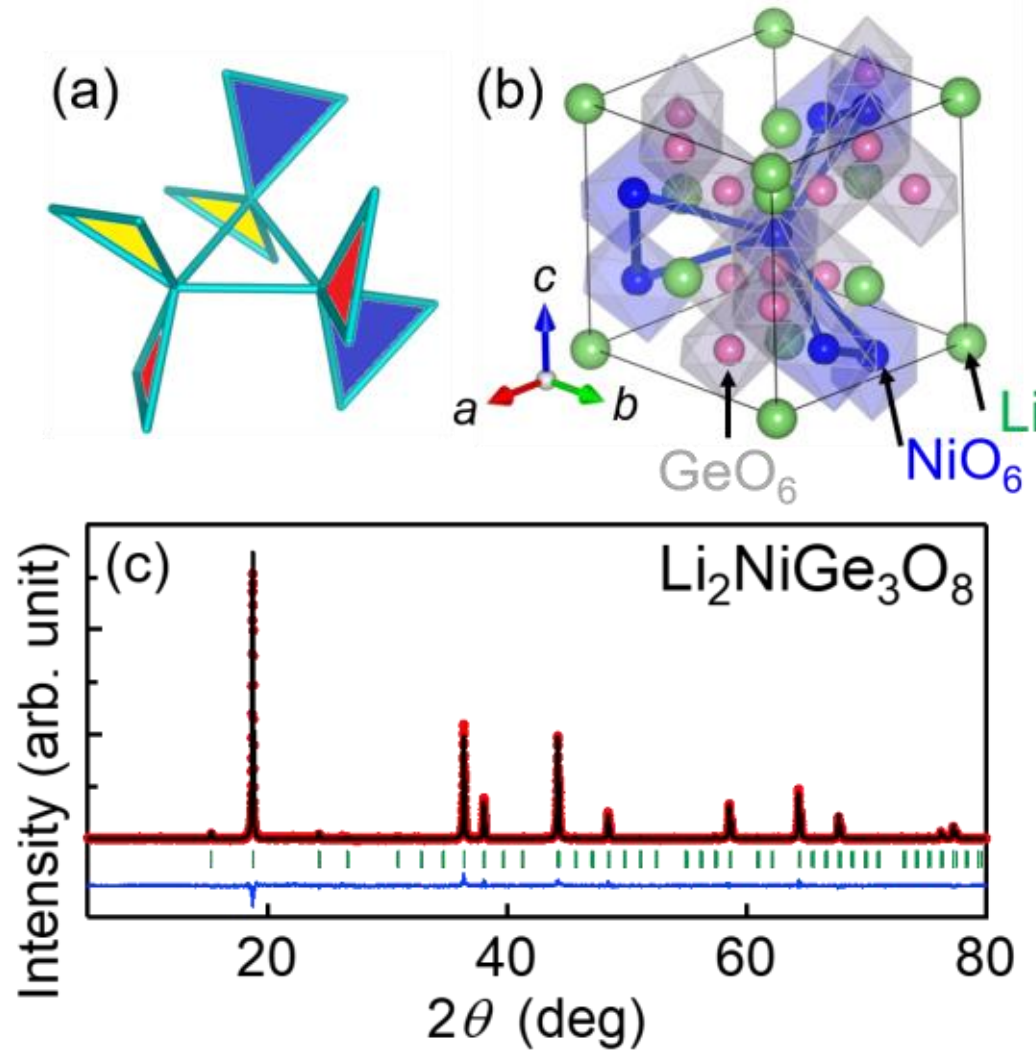

**Fig. 1** (a) Schematic view of the trillium lattice formed by the Ni sublattice in $Li_2NiGe_3O_8$. (b) Crystal structure of ordered-spinel $Li_2NiGe_3O_8$ highlighting the Ni network. (c) Powder XRD pattern and Rietveld refinement profile at room temperature. The crystallographic model is based on the reported ordered $Li_2NiGe_3O_8$ structure [42].

spin-ice model using a characteristic scale of order 7 K [18]. However, this comparison is used as a theoretical benchmark rather than as a definitive microscopic fit. In particular, the inverse susceptibility is compared only at a qualitative level, and the nearly vanishing Weiss temperature shows that $Li_2NiGe_3O_8$ cannot be described as a pure nearest-neighbour ferromagnetic Ising system.

We therefore adopt a deliberately conservative conclusion. $Li_2NiGe_3O_8$ is not established here as an ideal anisotropic trillium spin ice. Instead, it is presented as a rare $S$ = 1 single-trillium oxide with broad frustrated magnetic correlations, whose thermodynamics can be discussed in relation to both the Heisenberg-like and spin-ice-like limits of trillium-lattice magnetism. This position makes $Li_2NiGe_3O_8$ complementary to $K_2Ni_2(SO_4)_3$ and Na[Mn(HCOO)$_3$], while avoiding an overstatement of the present bulk data.

TABLE I. Crystallographic model of $Li_2NiGe_3O_8$ used for the Rietveld analysis. The fractional coordinates, Wyckoff sites, and occupancies are based on the ordered-spinel structure reported in Ref. [42]. Space group: $P4_132$ or $P4_332$ $a$ = 8.1799 Å. The coordinates and a single atomic displacement parameter $U_{iso}$ values are from Ref. [42] and were used as the ordered structural model for the present Rietveld analysis. The value $U_{iso}$ = 0.05 for Li should be regarded as part of the adopted structural model rather than as an independently refined parameter in the present work.

| Atom | Site | Occupancy | $x$ | $y$ | $z$ | $U_{iso}$ (Å$^2$) |
|---|---|---|---|---|---|---|
| Li | 8$c$ | 1 | 0.018(8) | = $x$ | = $x$ | 0.05 |
| Ni | 4$b$ | 1 | 5/8 | = $x$ | = $x$ | 0.012(4) |
| Ge | 12$d$ | 1 | 1/8 | 0.3777(7) | 0.8723(7) | 0.013(1) |
| O1 | 8$c$ | 1 | 0.385(3) | = $x$ | = $x$ | 0.01(1) |
| O2 | 24$e$ | 1 | 0.094(3) | 0.127(2) | 0.394(2) | 0.012(5) |

## II. Experimental Methods

Polycrystalline $Li_2NiGe_3O_8$ was synthesized by a conventional solid-state reaction. $LiOH·H_2O$, $Ni(OH)_2$, and $GeO_2$ powders were mixed in a nominal molar ratio of Li:Ni:Ge = 2.1:1:3, corresponding to a slight excess of $LiOH·H_2O$, pressed into pellets, and heated in air at 900 ºC for 24 h. The pellets were then reground, pelletized again, and annealed under the same conditions to improve homogeneity.

Powder x-ray diffraction (XRD) was measured at room temperature using Cu *K*α radiation. The diffraction pattern was analyzed by the Rietveld method using RIETAN-FP [40]. Structural refinement was performed assuming the ordered-spinel structure with space group $P4_132$ or $P4_332$, consistent with previous crystallographic work on $Li_2NiGe_3O_8$ and related $Li_2MGe_3O_8$ compounds [39,40,42]. The crystallographic model used for the Rietveld calculation was based on the reported ordered $Li_2NiGe_3O_8$ structure, in which Li occupies the tetrahedral site, Ni occupies the 4*b* octahedral site, and Ge occupies the 12*d* octahedral site [42].

Magnetization measurements were carried out in a Magnetic Property Measurement System (MPMS; Quantum Design) between 2 and 300 K under magnetic fields up to 7 T, at ISSP, the University of Tokyo. The temperature dependence of the dc susceptibility was measured using both zero-field-cooled and field-cooled protocols. Isothermal magnetization curves were also recorded at selected temperatures.

Heat capacity was measured in a Physical Property Measurement System (PPMS; Quantum Design) by the relaxation method, at ISSP, the University of Tokyo. Low-temperature data were collected in magnetic fields up to 10 T. To estimate the magnetic contribution, the lattice heat capacity was modeled by one Debye term and three Einstein terms fitted to the high-temperature data.

## III. Results

Figure 1 shows the crystal structure and powder XRD pattern of $Li_2NiGe_3O_8$. The powder XRD pattern is well reproduced by the ordered structural model summarized in Table I, with a lattice parameter a = 8.1799(2) Å taken from Ref. [42]. No impurity peaks are resolved within the present experimental sensitivity. The structure is consistent with space group $P4_132$ or $P4_332$ and agrees with earlier structural reports on $Li_2NiGe_3O_8$ and related $Li_2MGe_3O_8$ spinels [39,40,42].

Because the formation of the trillium Ni sublattice relies on Ni/Ge ordering, we explicitly describe the crystallographic model used for the Rietveld analysis. The structural parameters are listed in Table I. The fractional coordinates, isotropic displacement parameters, and site occupancies were taken from the ordered $Li_2NiGe_3O_8$ structure reported in Ref. [42]. In this ordered model, Li occupies the tetrahedral 8*c* site, Ni fully occupies the octahedral 4b site, and Ge fully occupies the octahedral 12d site. The Ni/Ge ordering is therefore explicitly included in the structural model used to construct the trillium Ni sublattice.

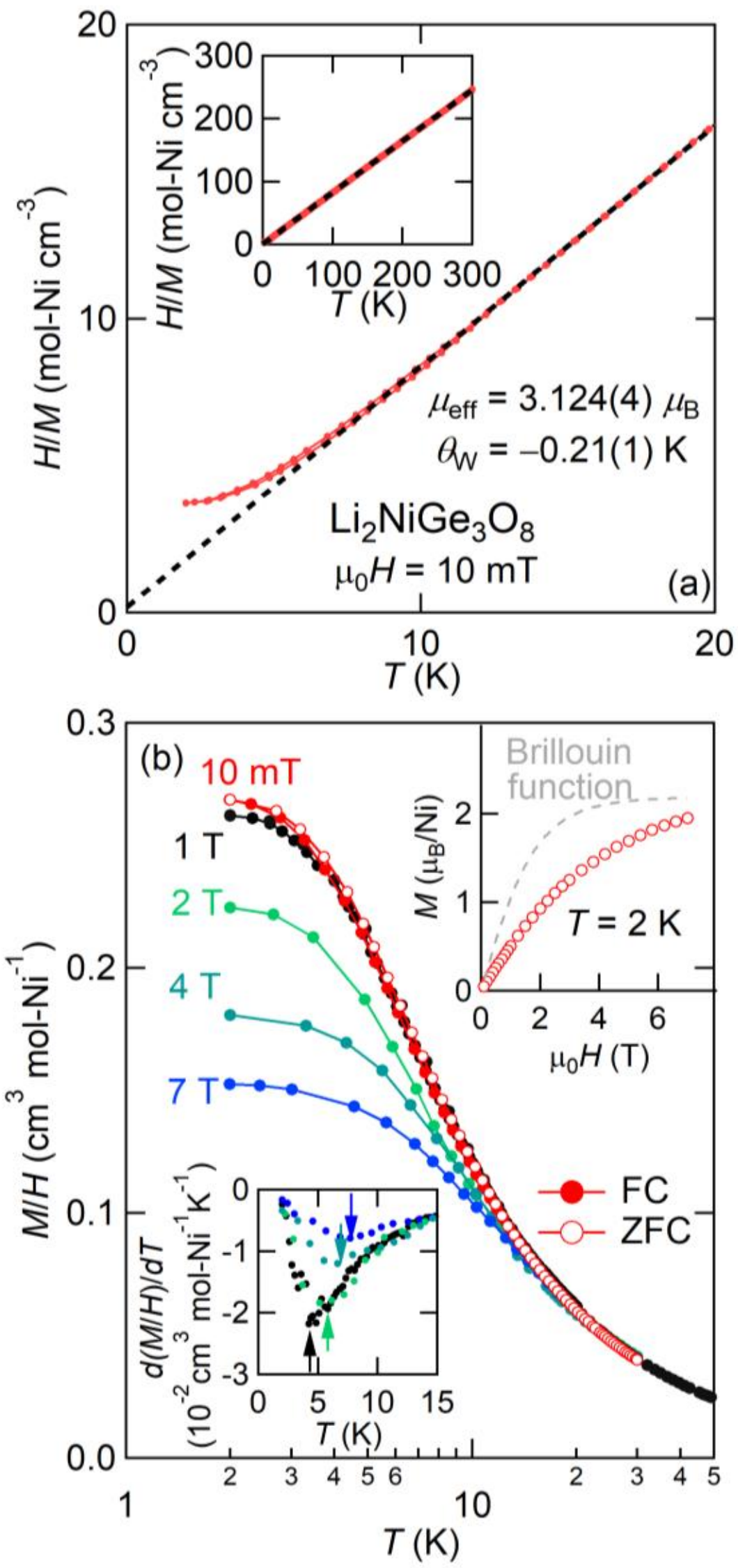

**Fig. 2** (a) Temperature dependence of the inverse magnetic susceptibility $H/M$ of $Li_2NiGe_3O_8$ measured at 0.01 T. The solid line is the Curie-Weiss fit. (b) Low-temperature susceptibility $M/H$ measured in several magnetic fields. Left inset: temperature derivative $d(M/H)/dT$ used to define the inflection temperature $T_{inf}$. Right inset: isothermal magnetization $M(H)$ at 2 K. The gray dashed line is the reference Brillouin function for isotropic noninteracting $S = 1$ moments with $g = 2.21$.

The present powder diffraction pattern is well reproduced by this ordered model without introducing additional cation disorder. We note, however, that the present laboratory powder XRD data are not sufficient to quantify a very small amount of Ni/Ge antisite disorder independently. Therefore, the magnetic discussion below is based on the ordered Ni-sublattice model supported by previous structural refinements and by the present Rietveld analysis, while the possible presence of a minor antisite component cannot be completely excluded from the present data alone.

The key structural feature for the present study is the Ni sublattice. In the ordered-spinel framework, $Ni^{2+}$ ions occupy octahedral 4*b* sites and their next-nearest-neighbour network forms the trillium lattice, namely a three-dimensional arrangement of corner-sharing equilateral triangles. Each Ni site belongs to three triangles, which makes this geometry intrinsically prone to frustration. Since the nonmagnetic Li and Ge ions separate the magnetic Ni sites, $Li_2NiGe_3O_8$ provides a relatively clean oxide platform for discussing the magnetism of a localized $S = 1$ trillium lattice

Figure 2(a) shows the inverse magnetic susceptibility $H/M$ measured at 0.01 T. Above 50 K, the data are well fitted by the Curie-Weiss law, which yields $\mu_{eff} = 3.124(4)\ \mu_B$ per Ni and $\theta_W = -0.21(1)$ K. The effective moment is close to the spin-only value expected for $S = 1$ and indicates that the magnetism is dominated by localized $Ni^{2+}$ moments. In contrast, the almost vanishing Weiss temperature implies that the net exchange field is very small on average, even though substantial correlations appear at much higher temperatures. This immediately suggests competition between ferromagnetic and antiferromagnetic interactions.

Below about 10 K, $H/M$ deviates smoothly from the high-temperature Curie-Weiss line. Such a deviation signals the buildup of short-range correlations well above the low-temperature anomaly in the heat capacity. This temperature scale is much larger than the absolute value of $\theta_W$, which again points to strong frustration or competing interactions.

Figure 2(b) shows the low-temperature susceptibility $M/H$ in several magnetic fields. $M/H$ increases on cooling and tends to saturate at low temperature. The derivative $d(M/H)/dT$ exhibits a minimum, which defines an inflection temperature $T_{inf}$. This characteristic temperature shifts to higher values with increasing magnetic field. In the following, $T_{inf}$ is used only as a field-dependent characteristic temperature and not as evidence by itself for a thermodynamic phase boundary. Zero-field-cooled and field-cooled data do not show appreciable hysteresis, which excludes simple spin-glass freezing in the measured temperature range. The isothermal magnetization at 2 K is nonlinear and bends gradually with increasing field. Moreover, the $M(H)$ curve is not adequately described by a simple Brillouin function for noninteracting isotropic $S = 1$ moments. The gray dashed line in the right inset represents the corresponding reference Brillouin function with $g = 2.21$ estimated from the Curie-Weiss analysis and is shown only as an illustrative guide. Although this deviation does not by itself uniquely prove short-range order, because single-ion anisotropy and powder averaging can also modify the field dependence, it does show that the low-temperature state cannot be understood as a trivial free-spin paramagnet.

Figure 3(a) shows the temperature dependence of $C/T$. To extract the magnetic contribution, the lattice term $C_{latt}$ was fitted by a Debye contribution plus three Einstein contributions,

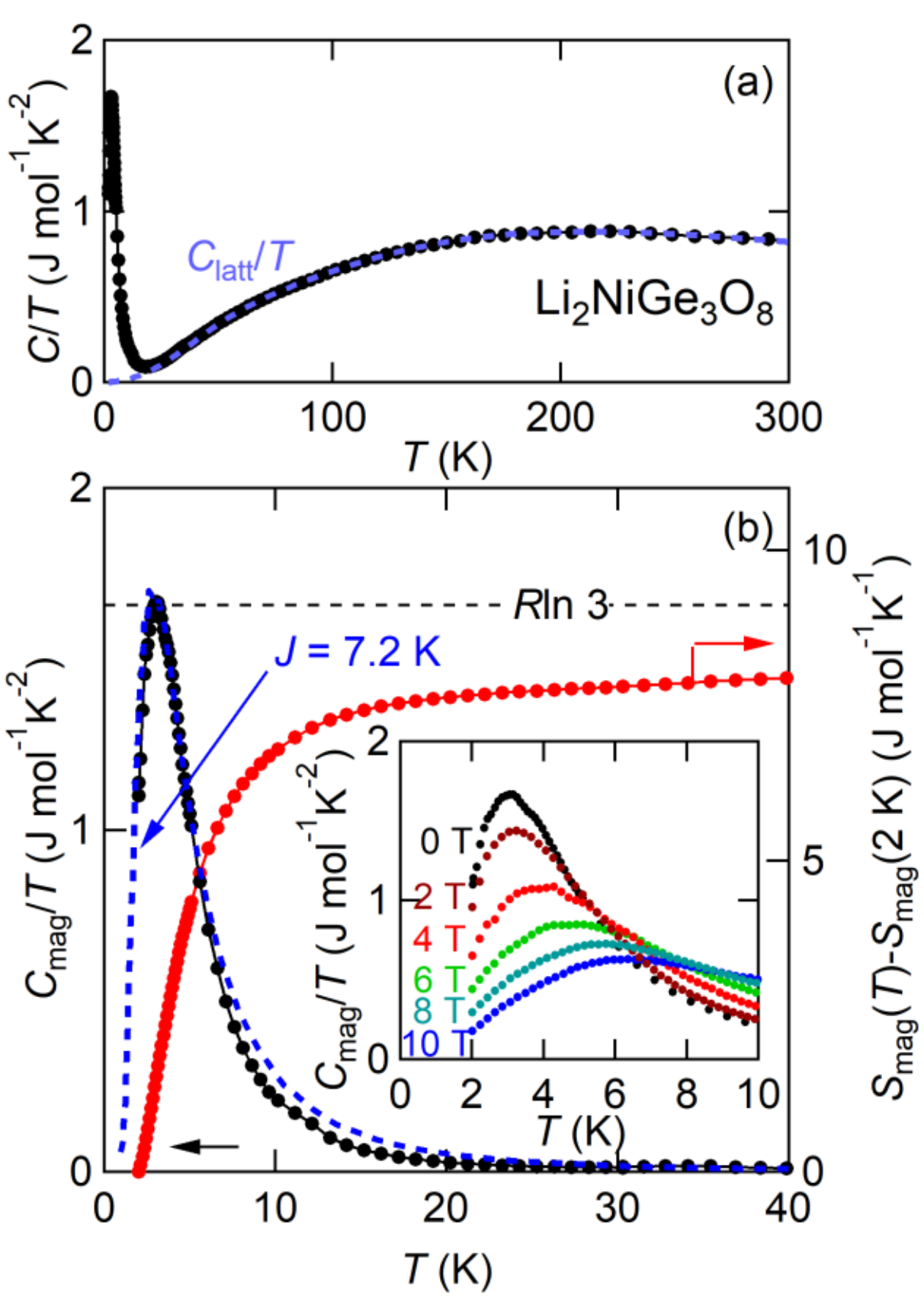

**Fig. 3** (a) Temperature dependence of $C/T$ for $Li_2NiGe_3O_8$. The dashed line shows the fitted lattice contribution. (b) Magnetic heat capacity $C_{mag}/T$ and magnetic entropy $S_{mag}$. The dashed curve represents the Monte Carlo heat-capacity curve for the local ferromagnetic Ising model on the trillium lattice, scaled with an effective characteristic energy $J$ of order 7 K. This comparison is used as a theoretical benchmark, not as a complete microscopic fit. Inset: field dependence of $C_{mag}/T$.

$$C_{latt} = C_D + C_E = 9R(T/\theta_D)^3 \int_0^{\theta_D/T} \frac{x^4 \exp(x)}{[\exp(x)-1]^2} dx + R\sum_{i=1}^{3} n_i \frac{(\theta_{Ei}/T)^2 \exp(\frac{\theta_{Ei}}{T})}{\left[\exp(\frac{\theta_{Ei}}{T})-1\right]^2}$$

Here, $R$ is the gas constant, $\theta_D$ is the Debye temperature, $\theta_{Ei}$ are the Einstein temperatures with degeneracies $n_i$, and sum from $i = 1$ to 3 of $n_i = 3n - 3 = 39$. The best fit is shown by the dashed line with $\theta_D = 327$ K, $\theta_{E1} = 317$ K, $\theta_{E2} = 679$ K, $\theta_{E3} = 1772$ K, $n_1 = 3$, $n_2 = 28$, and $n_3 = 8$. The fitted curve reproduces the high-temperature specific heat well and serves as the basis for estimating $C_{mag}$. In the temperature range of the broad magnetic anomaly, the lattice contribution is already small, so the existence of the broad feature discussed below is robust against reasonable variations in the Debye-Einstein fit.

Figure 3(b) shows $C_{mag}/T$ below 40 K together with the magnetic entropy $S_{mag}$ obtained by integrating $C_{mag}/T$ with respect to temperature. The most important experimental fact is that $C_{mag}/T$ exhibits a broad maximum centered at about 3 K rather than a sharp lambda-type peak. A broad magnetic tail persists to about 10 K, which means that the entropy is released over a wide temperature range. The recovered entropy between 2 and 40 K is about 8 J·mol$^{-1}$K$^{-1}$, corresponding to approximately 88% of $R$ ln 3.

Because the measurement extends only down to 2 K and because the lattice subtraction carries some uncertainty, this value should not be interpreted too aggressively as a quantitative residual entropy. The reliable statement is simply that the magnetic entropy is released gradually over a broad temperature interval and is not exhausted at a single narrow anomaly. The present data also do not exclude a distinct low-temperature transition below 2 K.

The inset of Fig. 3(b) shows the field dependence of $C_{mag}/T$ up to 10 T. As in the case of $T_{inf}$, the peak positions are treated here as field-dependent characteristic temperatures or crossover-like scales rather than as a phase boundary inferred from the present bulk data alone. This field evolution tracks the behavior of $T_{inf}$ extracted from $M/H$ and will later be compared with the theoretical expectations for the trillium spin-ice model.

## IV. Discussion

Before discussing specific models, it is important to separate robust experimental facts from interpretation. The robust facts are as follows. First, $Li_2NiGe_3O_8$ is a single-trillium oxide with localized $S$ = 1 $Ni^{2+}$ moments. Second, the Weiss temperature is almost zero, while magnetic correlations develop on the scale of 10 K. Third, the magnetic heat capacity does not show a sharp transition-like lambda anomaly near 3 K, but rather a broad maximum with a wide tail. Fourth, the characteristic temperatures extracted from susceptibility and heat capacity both shift upward with field. In the present work, these temperatures are regarded as crossover-like characteristic scales rather than as a thermodynamic phase boundary inferred from bulk data alone.

A broad magnetic heat-capacity maximum and a gradual entropy release are not unique to spin-ice physics. They are commonly observed in geometrically frustrated magnets, where short-range correlations develop over a wide temperature range above any eventual ordering or freezing transition. Therefore, the broad $C_{mag}/T$ maximum in $Li_2NiGe_3O_8$ should not by itself be regarded as evidence for an anisotropic spin-ice state. The more conservative conclusion from the present data is that $Li_2NiGe_3O_8$ hosts broad frustrated magnetic correlations on an $S$ = 1 trillium lattice. The trillium spin-ice model is used below as a theoretically established limiting model on the same lattice, not as a complete microscopic description of the material.

The most suggestive theoretical framework for the present data is the local ferromagnetic Ising model on the trillium lattice studied by Redpath and Hopkinson [18]. In that model, the most characteristic thermodynamic signature is a broad soft peak in the heat capacity rather than a sharp ordering anomaly [18]. This aspect is directly relevant to $Li_2NiGe_3O_8$, because the magnetic heat capacity shows a broad maximum near 3 K with a wide tail extending to about 10 K, rather than a narrow lambda-type transition. Within the present level of analysis, the comparison of $C_{mag}/T$ with the Monte Carlo result of Ref. [18] provides the primary basis for introducing a characteristic energy scale of order 7 K.

By contrast, the inverse susceptibility should be compared more cautiously. In the Monte Carlo study of Ref. [18], the inverse susceptibility is nearly linear at high temperature, then turns upward relative to the Curie-Weiss line below about 3$J$, and develops only a weak shoulder before approaching zero at low temperature. $Li_2NiGe_3O_8$ does show a similar qualitative evolution in the limited sense that $H/M$ is Curie-Weiss-like at high temperature and deviates smoothly from the straight line below about 10 K. However, the experimental curve is not quantitatively reproduced by the Monte Carlo result of Ref. [18], and we do not claim that the same 7 K scale provides a direct fit to the inverse susceptibility.

This distinction is physically important. In the ideal trillium spin-ice model, the high-temperature inverse susceptibility corresponds to a positive Curie temperature of order $J$ [18], whereas in $Li_2NiGe_3O_8$ the Weiss temperature is nearly zero and slightly negative. The real material is therefore substantially displaced from the pure nearest-neighbour local ferromagnetic Ising limit. What is shared between experiment and theory is not the detailed curve shape, but the broader thermodynamic fingerprint: a smooth low-temperature departure from Curie-Weiss behavior together with a broad heat-capacity maximum. We therefore regard the heat-capacity comparison as semi-quantitative, whereas the inverse-susceptibility comparison remains qualitative.

The nearly vanishing $\theta_W$ also indicates that a single ferromagnetic Ising interaction cannot be the only relevant interaction in $Li_2NiGe_3O_8$. If the dominant interaction were a simple nearest-neighbour ferromagnetic Ising coupling, a positive Curie-Weiss temperature would be expected. The observed $\theta_W$ near zero instead suggests that ferromagnetic and antiferromagnetic contributions nearly compensate in the high-temperature uniform susceptibility. Thus, the scale $J$ of order 7 K inferred from the heat-capacity comparison should be regarded as an effective characteristic scale of the broad correlated regime, not as a unique microscopic exchange constant. In a real material with several interactions of different signs and comparable magnitudes, the ideal spin-ice model can serve only as a limiting reference model.

It is also important to ask how far $Li_2NiGe_3O_8$ is from an isotropic Heisenberg magnet. The present powder bulk measurements do not allow us to determine the magnitude of the magnetic anisotropy quantitatively. Therefore, we do not claim that $Li_2NiGe_3O_8$ is an established anisotropy-dominated $S = 1$ trillium magnet. Broad heat-capacity maxima can also be produced in frustrated Heisenberg systems as a consequence of short-range correlations. Indeed, theoretical studies of the Heisenberg antiferromagnet on the trillium lattice and the experimental realization $Na[Mn(HCOO)_3]$ demonstrate that Heisenberg-type trillium magnets can show broad correlated regimes above low-temperature ordering [15-17,30]. The distinction in the present work is more modest: $Li_2NiGe_3O_8$ shows a broad heat-capacity maximum whose scale and shape can be compared with the soft peak of the trillium spin-ice model, while $\theta_W$ remains nearly zero, indicating competing interactions. Thus, the current data place $Li_2NiGe_3O_8$ in the broader landscape between Heisenberg-like and spin-ice-like limits, rather than proving that it belongs to the latter.

At this stage, a central question is whether $Ni^{2+}$ in $Li_2NiGe_3O_8$ can support the anisotropy required for proximity to the trillium spin-ice regime. In the ideal cubic limit, the lowest-order single-ion cubic anisotropy of $S = 1$ $Ni^{2+}$ vanishes, so $Ni^{2+}$ is not generically Ising-like. However, this does not imply that $Ni^{2+}$ is strictly isotropic. Yosida and Tachiki showed for Ni-ferrite that higher-order anisotropy can survive through the combined action of spin-orbit coupling and exchange interactions: when exchange is included in the unperturbed manifold, the anisotropy appears already in the fourth-order perturbed energy, and the same contribution is recovered in sixth order when exchange is treated perturbatively together with the $L\cdot S$ coupling [38]. Their result provides an important proof of principle that $Ni^{2+}$ in an approximately cubic environment can acquire an effective easy-axis anisotropy through exchange-assisted spin-orbit processes. In this sense, the relevant anisotropy in $Li_2NiGe_3O_8$ need not arise solely from a local single-ion term, but may also contain an exchange-assisted contribution of the Yosida-Tachiki type mechanism [38]. At present, however, the microscopic origin and magnitude of this anisotropy remain open questions.

This point also clarifies why $K_2Ni_2(SO_4)_3$ does not make the present material redundant, even though both compounds contain $S = 1$ $Ni^{2+}$. The essential difference is not the magnetic ion but the effective Hamiltonian and lattice topology. $K_2Ni_2(SO_4)_3$ is a system of two coupled trillium lattices whose exchange network has been described on the Heisenberg side, with weak bulk anisotropy and field-induced quantum-spin-liquid behavior [28,29]. By contrast, the Redpath-Hopfkinson model is a single-trillium local-axis ferromagnetic Ising model [18]. $Li_2NiGe_3O_8$ is interesting because it offers a different single-trillium oxide platform in which the broad thermodynamic response can be compared with the spin-ice-like limiting model, even though the present data do not quantify the anisotropy.

A professional reading of the data should also consider alternatives. A frustrated Heisenberg antiferromagnet with moderate single-ion anisotropy, weak canting, or further-neighbour exchange could in principle produce a broad heat-capacity maximum and a nontrivial low-temperature magnetization [15-17,30-33]. Likewise, a partially ordered or cluster-correlated state could broaden the thermodynamic anomaly without invoking a true spin-ice manifold. Bulk measurements alone do not rule out these possibilities.

For this reason, our interpretation is intentionally limited. We do not conclude that $Li_2NiGe_3O_8$ realizes an anisotropic trillium spin ice. Rather, we conclude that the bulk thermodynamics of $Li_2NiGe_3O_8$ show broad frustrated magnetic correlations that are usefully compared with the spin-ice-like limiting model on the same trillium lattice. The heat-capacity comparison gives an effective scale of order 7 K, while the inverse susceptibility provides only qualitative support through its smooth low-temperature departure from Curie-Weiss behavior. Further microscopic probes are required to decide whether the underlying correlations are closer to an anisotropic local-axis model, an anisotropic Heisenberg model, or another frustrated correlated state.

The present data also leave open a second important point: a distinct transition below 2 K cannot be excluded. This uncertainty further supports a cautious formulation of the claim. What the present measurements establish is not the exact ground state, but the character of the broad correlated regime above 2 K.

This point clarifies the broader impact of the work. $K_2Ni_2(SO_4)_3$ realizes two coupled $S = 1$ trillium lattices and shows field-induced quantum spin-liquid behavior [28,29]. $Na[Mn(HCOO)_3]$ realizes the nearest-neighbour Heisenberg antiferromagnet on a single trillium lattice and displays classical spin-liquid behavior above its ordering temperature [30]. $KSrFe_2(PO_4)_3$, $K_2CrTi(PO_4)_3$, and $KBaCr_2(PO_4)_3$ show that the family already extends to additional frustrated phosphate-based trillium materials with broad correlated regimes, persistent dynamics, or multiple anomalies [31-33]. $Li_2NiGe_3O_8$ adds a rare oxide example of a single $S = 1$ trillium lattice. Its significance is that it provides a bulk-thermodynamic platform for comparing Heisenberg-like and spin-ice-like frustrated regimes on the same lattice geometry.

The present work strongly motivates further experiments. Heat-capacity measurements below 2 K are needed before any quantitative statement about residual entropy can be made. Single-crystal measurements would be highly valuable for determining magnetic anisotropy directly and testing whether local easy axes are present, especially in view of the sensitivity of $Ni^{2+}$ anisotropy to local coordination geometry [34-37]. Neutron diffraction or muon spin relaxation is required to determine whether a truly static ordered state develops below the broad maximum in $C_{mag}/T$. Inelastic neutron scattering or electron spin resonance would help establish whether an effective Ising description is microscopically justified. These experiments will determine how close $Li_2NiGe_3O_8$ lies to the trillium spin-ice limit and what kind of low-temperature state ultimately emerges.

## V. Summary

In summary, we have investigated the ordered-spinel oxide $Li_2NiGe_3O_8$, in which localized $S$ = 1 $Ni^{2+}$ moments form a single three-dimensional trillium lattice, by means of powder XRD, magnetization, and heat-capacity measurements. The inverse susceptibility $H/M$ follows Curie-Weiss behavior above 50 K with $\mu_{eff}$ = 3.124(4) $\mu_B$ per Ni and $\theta_W$ = –0.21(1) K, but deviates smoothly from the high-temperature line below about 10 K, indicating the development of short-range correlations on an energy scale much larger than the mean Weiss field. The magnetic heat capacity $C_{mag}/T$ exhibits a broad maximum near 3 K with a wide tail extending to about 10 K, and the entropy recovered between 2 and 40 K reaches about 88% of $R$ ln 3, showing that the magnetic entropy is released gradually rather than at a sharp lambda-type anomaly. A semi-quantitative comparison of $C_{mag}/T$ with the Monte Carlo result for the local ferromagnetic Ising model on the trillium lattice suggests a characteristic scale $J$ of order 7 K, while the inverse susceptibility is similar to the theoretical result only at a qualitative level. The present data do not establish $Li_2NiGe_3O_8$ as an ideal anisotropic trillium spin ice, nor do they quantify the magnetic anisotropy. Rather, they establish $Li_2NiGe_3O_8$ as a rare single-trillium oxide with broad frustrated magnetic correlations and motivate further microscopic studies of its relation to the Heisenberg-like and spin-ice-like limits of trillium-lattice magnetism.

## Acknowledgement

This work was supported by JST PRESTO Grant Number JPMJPR23Q8 (Creation of Future Materials by Expanding Materials Exploration Space) and JSPS KAKENHI Grant Numbers. JP25K01496 (Scientific Research (B)), JP23H04616 and JP25H01649 (Transformative Research Areas (A) "Supra-ceramics"), JP25H01403 (Transformative Research Areas (B) "Multiply Programmed Layers"), and JP22K14002 (Young Scientific Research) Part of this work was carried out by joint research in the Institute for Solid State Physics, the University of Tokyo (Project Numbers 202311-MCBXG-0021, 202311-MCBXG-0025, 202406-MCBXG-0100, 202406-GNBXX-0095, 202406-MCBXG-0100, 202406-MCBXG-0101, 202411-MCBXG-0033, and 202411-MCBXG-0034).